\documentclass[prl,aps,amsmath,amssymb,reprint]{revtex4-1}
\usepackage{graphicx}
\usepackage{dcolumn}
\usepackage{bm}
\newcommand{\p}{$\%$}

\begin{document}
\title{Density and Microstructure of $a$-C Thin Films}
\author{Prabhat Kumar}\address{UGC-DAE
Consortium for Scientific Research, University Campus, Khandwa
Road, Indore 452 001, India}
\author{Mukul Gupta}\address{UGC-DAE
Consortium for Scientific Research, University Campus, Khandwa
Road, Indore 452 001,
India}\email{$^{\dag}$mgupta@csr.res.in/dr.mukul.gupta@gmail.com}
\author{Jochen Stahn} \address{Laboratory for Neutron Scattering and Imaging, Paul
Scherrer Institut, CH-5232 Villigen PSI, Switzerland}
\author{U. P. Deshpande}\address{UGC-DAE
Consortium for Scientific Research, University Campus, Khandwa
Road, Indore 452 001, India}
\author{D. M. Phase}\address{UGC-DAE
Consortium for Scientific Research, University Campus, Khandwa
Road, Indore 452 001, India}
\author{V. Ganesan}\address{UGC-DAE
Consortium for Scientific Research, University Campus, Khandwa
Road, Indore 452 001, India}
\date{\today}


\begin{abstract}

In this work, we studied amorphous carbon ($a$-C) thin films
deposited using direct current (dc) and high power impulse
magnetron sputtering (HiPIMS) techniques. The microstructure and
electronic properties reveal subtle differences in $a$-C thin
films deposited by two techniques. While, films deposited with
dcMS have a smooth texture typically found in $a$-C thin films,
those deposited with HiPIMS consist of dense hillocks surrounded
by a porous microstructure. The density of $a$-C thin films is a
decisive parameter to judge their quality. Often, x-ray
reflectivity (XRR) has been used to measure the density of carbon
thin films. From the present work, we find that determination of
density of carbon thin films, specially those with a thickness of
few tens of nm, may not be accurate with XRR due to a poor
scattering contrast between the film and substrate. By utilizing
neutron reflectivity (NR) in the time of flight mode, a technique
not commonly used for carbon thin films, we could accurately
measure differences in the densities of $a$-C thin films deposited
using dcMS and HiPIMS.

\end{abstract}

\maketitle

\section{Introduction}\label{1}
Amorphous carbon ($a$-C), graphite and diamond are well-known
allotropes of carbon. Graphite with a layered hexagonal crystal
structure has sp$^{2}$ hybridization and in diamond due to a
tetrahedral structure, hybridization is sp$^{3}$. On the other
hand, $a$-C is somewhere in between, having a mixture of both
hybridizations~\cite{DLCROBERTSON2002129}. Due to presence of a
significant fraction of sp$^{3}$ hybridization and properties
similar to diamond, ($a$-C) is also referred as diamond like
carbon (DLC). $a$-C has several interesting properties $e.g.$ they
are mechanically hard, chemically inert and transparent (mainly in
infrared region)~\cite{DLCROBERTSON2002129}, etc. Therefore, $a$-C
films are widely used as protective coatings of magnetic data
storage devices, biomedical equipments, optical windows and cold
neutron applications. With such an excellent physical, mechanical
and chemical properties, it has gained attention among the
researcher and in coating technology.

An $a$-C thin film can be deposited using chemical vapor
deposition (CVD) or physical vapor deposition (PVD) ($e.g.$ pulse
laser deposition (PLD), sputtering) techniques. When deposited
using CVD, generally $a$-C thin films have hydrogen(H)
contamination. Among the PVD techniques, PLD has been frequently
used to deposit $a$-C films having sp$^{3}$ fraction as high as
80\,\p~\cite{DLCPLDJJGAUMET1993, DLCPLDYAMAMOTO1998}. On the other
hand, it is well-known that small sample size and lower deposition
rates limit the uses of PLD for mass production and industrial
uses. Another PVD technique that is magnetron sputtering (MS), has
also emerged as an industrially accepted technique for preparation
of H free $a$-C thin films. Though the sp$^{3}$ fraction is found
to be typically 45\,\p~ in sputtered
film~\cite{DLCLOGOTHETIDIS1999SCT}. It may be noted that in a
typical direct current MS (dcMS) process, the plasma is dominated
by neutrals and the fraction of ions is very small ($<$5\p). On
the onset of this century, an advancement in dcMS was seen in
terms of high power impulse magnetron sputtering (HiPIMS). In
HiPIMS very large power (about 10$^{3}$ greater than that in dcMS)
may enhance the fraction of ions to the extent that it can be even
larger than neutral atoms. HiPIMS technique was immediately
applied for preparation of thin films of transition metal
compounds e.g. TiO$_{2}$~\cite{TiOHiPIMSKONSTANTINIDIS20061182},
TiN~\cite{TiNHiPIMSLATTEMANN20105978}, CrN~\cite{CrNHiPIMSLIN2011}
etc. A general observation in terms of properties of deposited
films was observed as the morphology was globular rather than
columnar observed in dcMS. Sporadic attempts have also been made
to deposit $a$-C films with HiPIMS. However, in most of the
studies so far, the characterization of plasma has been the focus,
rather than the properties of resulting
films~\cite{LATTEMANN2011DRMHIPIMS,
SARAKINOS2012SuCoTechDLCHIPIMS, NAKAO2013IEEEDLCHIPIMS,
HUANG2013ASSDLCHIPIMS, LIN2014SuCoTechDLCHIPIMS}. Lattemann $et
al.$~\cite{LATTEMANN2011DRMHIPIMS} used HiPIMS together with arc
mode to find enhancement in C ions and resulting films were found
to have graphitic clusters. In another study with rather high
frequency (250\,Hz to 1\,kHz), the plasma was found to have larger
ionized species yielding enhance densification with larger
sp$^{3}$ fraction~\cite{SARAKINOS2012SuCoTechDLCHIPIMS}. In more
recent, works Ar and C$_{2}$H$_{2}$ gas mixture were used to
deposit $a$-C films but films were not H
free~\cite{DLCHIPIMSTAKASHI2016,DLCHIPIMSASIMJVS2016}. Clearly,
HiPIMS offers possibilities to deposit carbon thin films with high
density and high sp$^{3}$ fractions, and more systematic studies
are required to achieve this.

Microstructure and density of $a$-C thin films determines their
merit. The microstructure and the electronic structure of $a$-C
thin films have been well-established and techniques employed to
measure them are: transmission electron microscopy (TEM), atomic
force microscopy (AFM), scanning electron microscopy (SEM); and
electronic structure using Raman, x-ray photo electron
spectroscopy (XPS) and C K-edge x-ray absorption spectroscopy
(XAS). In detailed study by Ferrari et
al.~\cite{FERRARI2000PRBDENSITY} it was found that there is near
linear dependence of density with sp$^{3}$ fraction. Therefore
density of $a$-C thin films is pivotal to determine their quality.
Generally, x-ray reflectivity (XRR) has been the technique used
most frequently to measure the density of $a$-C thin films. In
most of the studies an $a$-C film has been deposited on a Si
substrate. The density of $a$-C (typically 2.2\,g/cm$^{3}$) films
is slightly less than that of Si (2.3\,g/cm$^{3}$). In a typical
XRR pattern two critical angles are
expected~\cite{FERRARI2000PRBDENSITY}. However, their separation
is too close to observe experimentally and it may happen that the
critical angle of $a$-C film get merged together with Si. In this
case, rather than measuring the density of $a$-C film, the density
of Si substrate is measured. It may be noted that in recent works
though density of $a$-C films have been determined using
XRR~\cite{LATTEMANN2011DRMHIPIMS, SARAKINOS2012SuCoTechDLCHIPIMS},
either XRR patterns were not shown or the separation of critical
edges of C and Si could not be seen~\cite{LIU2016DRM}. Moreover,
as pointed out in a seminal work by Wallace et
al.~\cite{WALLACE1995APL}, determination of density from typical
XRR measurements (angle dispersive $\theta$ - 2\,$\theta$ scans)
could have errors up to 5\,\p~due to sample misalignment and a
similar error may get augmented as the critical edge is not well
defined due to similarity in densities of C and Si.

It is surprising to note that neutron reflectivity (NR), which is
otherwise similar to XRR, has not yet been used to determine the
density of $a$-C films. For C and Si, neutron scattering length
densities are: 7.33$\times$10$^{-6}$ and
2.04$\times$10$^{-6}$\,{\AA$^{-2}$}; for Cu K$\alpha$ x-rays they
are: 1.87$\times$10$^{-5}$ and 2.01$\times$10$^{-5}$ {\AA$^{-2}$},
respectively. Since for neutrons, C is a stronger scatterer,
therefore the critical angle is exclusively determined by C.
Moreover, by doing NR measurements in energy dispersive or time of
flight (TOF) mode, the angle of incidence is kept fixed (no
movement of sample during measurement). Therefore foot print
effects can be avoided completely and the critical angle of the
density of C thin films can be measured much more precisely than
XRR. In the present work, we have shown this amply for C thin
films deposited using dcMS and HiPIMS techniques. As shown in this
work, with XRR small difference in the density of C films could
not be probed, they could be seen clearly with NR. In addition
synchrotron based C K-edge x-ray absorption spectroscopy (XAS) was
used to investigate the nature of bonding and hybridization.
Combining XRR, NR and XAS data with laboratory based techniques -
AFM and XPS, we probed the microstructure and density of $a$-C
thin films.

\section{Experimental Details}\label{2}
Carbon thin films were deposited using dcMS and HiPIMS at room
temperature (without any intentional heating) using a 3\,inch C
(99.999\% purity) target (Orion-8, AJA Int. Inc. system). The base
pressure of the chamber was of the order of 1$\times10^{-7}$\,Torr
and working pressure was 3\,mTorr due to flow of Ar gas (99.9995\%
purity) at 20\,sccm. It is expected that by using a pure C target,
low background pressure and pure Ar gas, the resulting films would
be hydrogen free, unlike those obtained by a chemical
precoces~\cite{DLC:HWEILER19942797}. Films were deposited at an
average power (current) of 100\,W (0.2\,A), both for dcMS and
HiPIMS process but the peak power (current) obtained in the later
was about 28\,kW (56\,A). Typical deposition times were about 1\,h
in dcMS and 2\,h in the HiPIMS process. The duty cycle used in the
HiPIMS process was about 0.35~\p~(pulse length 70\,$\mu$\,s and
frequency 50\,Hz). Samples were deposited on a Si(100) substrate
and the target to substrate distance was kept fixed at about
12\,cm. For better uniformity substrates were continuously rotated
around their own axis at 60\,rpm. Resulting films were
characterized for their thickness, density and roughnesses by
x-ray reflectivity (XRR) using a standard diffractometer (Bruker
D8 Discover) equipped with a Cu k$_\alpha$ x-ray source. To gain
further insight about parameters obtained from XRR measurements,
we did neutron reflectivity (NR) measurements at AMOR
reflectometer in time of flight mode at SINQ/PSI,
Switzerland~\cite{AMORSTAHN201644, AMORGupta2004}. The electronic
structure of the samples were determined using x-ray absorption
near edge spectroscopy (XANES) technique at the C K-edge in total
electron yield (TEY) mode at BL01 beamline at Indus-2 synchrotron
radiation source at RRCAT, Indore~\cite{BL01}. We have also used
x-ray photoelectron spectroscopy (XPS) to study the electronic
structure. The XPS measurement were carried out using Electron
Spectroscopy for Chemical Analysis (ESCA) spectrometer equipped
with Al k$_\alpha$ x-ray source. Both XANES and XPS measurements
were performed under UHV conditions. Surface morphology of the
deposited samples were determined by atomic force microscopy(AFM)
operating in contact mode.

\section{Results and Discussion} \label{3}
\subsection{Atomic Force Microscopy} \label{3.1}
\begin{figure} \center \vspace{5mm}
\includegraphics [width=90mm,height=75mm] {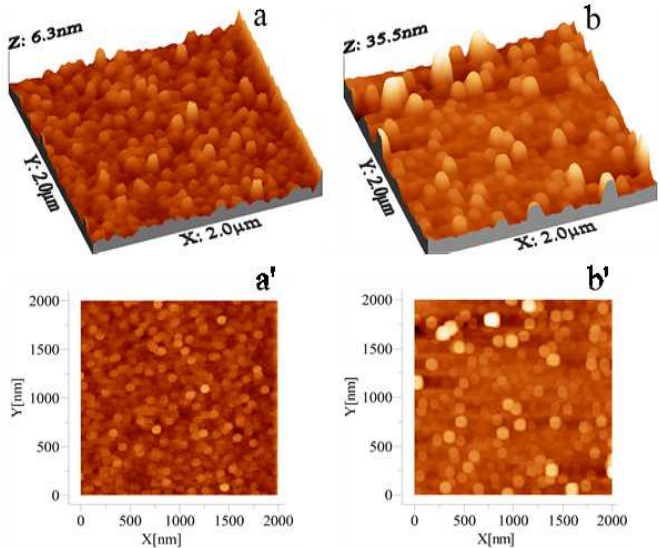} \vspace{-2mm}
\caption{\label{afm} (Color online) 3D view of AFM of carbon thin
film deposited by dcMS (a) and HiPIMS (b) The X and Y scale is
2$\times$2 $\mu$m$^{2}$. Tob view of the AFM images a$^\prime$
(dcMS) and b$^\prime$ (HiPIMS) with z-scale same as in 3D view.}
\vspace{-1mm}
\end{figure}

Surface morphologies of the deposited samples obtained from AFM
measurements are shown in fig.~\ref{afm}. Images were processed
using WSxM software package~\cite{WSXM2007}. Keeping X and Y scale
constant in all images at 2$\times$2 $\mu$m$^{2}$, the Z-scale was
varied on the basis of maximum column height. It is about
65\,\AA~for films grown with dcMS (fig.~\ref{afm} (a),
(a$^\prime$)) and about 350\,\AA~for samples grown with
HiPIMS(fig.~\ref{afm} (b), (b$^\prime$). As can be seen from AFM
images, film grown using HiPIMS has topographical distribution
which is about 5 times higher in height than those in dcMS
deposited films. This indicates that while films deposited using
dcMS are smooth, those deposited using HiPIMS have a distribution
in which large columns or hillocks are surrounded by voids. It is
known that the density of ions is larger in the HiPIMS process and
when deposited film gets bombarded with ions, this could result in
such distribution leading to formation of a porus microstructure.
The consequences of these are also reflected in the density as
well as in the orbital ordering, which is measured with XRR, NR,
XPS and C K-edge XANES and presented in next section.

\begin{figure} \center
\vspace{-1mm}
\includegraphics [width=85mm,height=65mm] {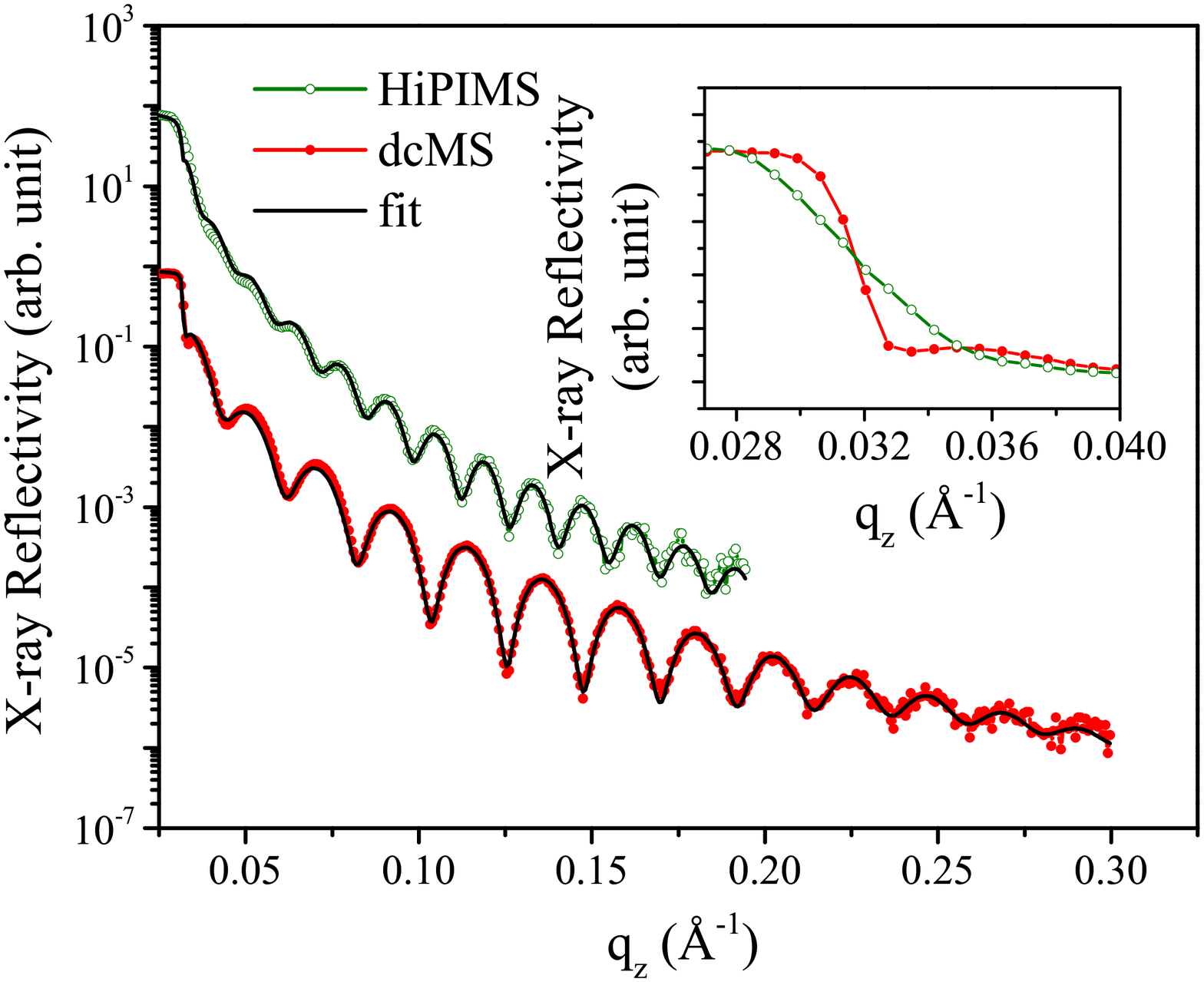} \vspace{-2mm}
\caption{\label{CXRR} (Color online) X-ray reflectivity of the
deposited carbon thin films using dcMS and HiPIMS, the inset shows
the closer view of the critical angles.} \vspace{-1mm}
\end{figure}

\begin{figure} \center
\vspace{-1mm}
\includegraphics [width=85mm,height=70mm] {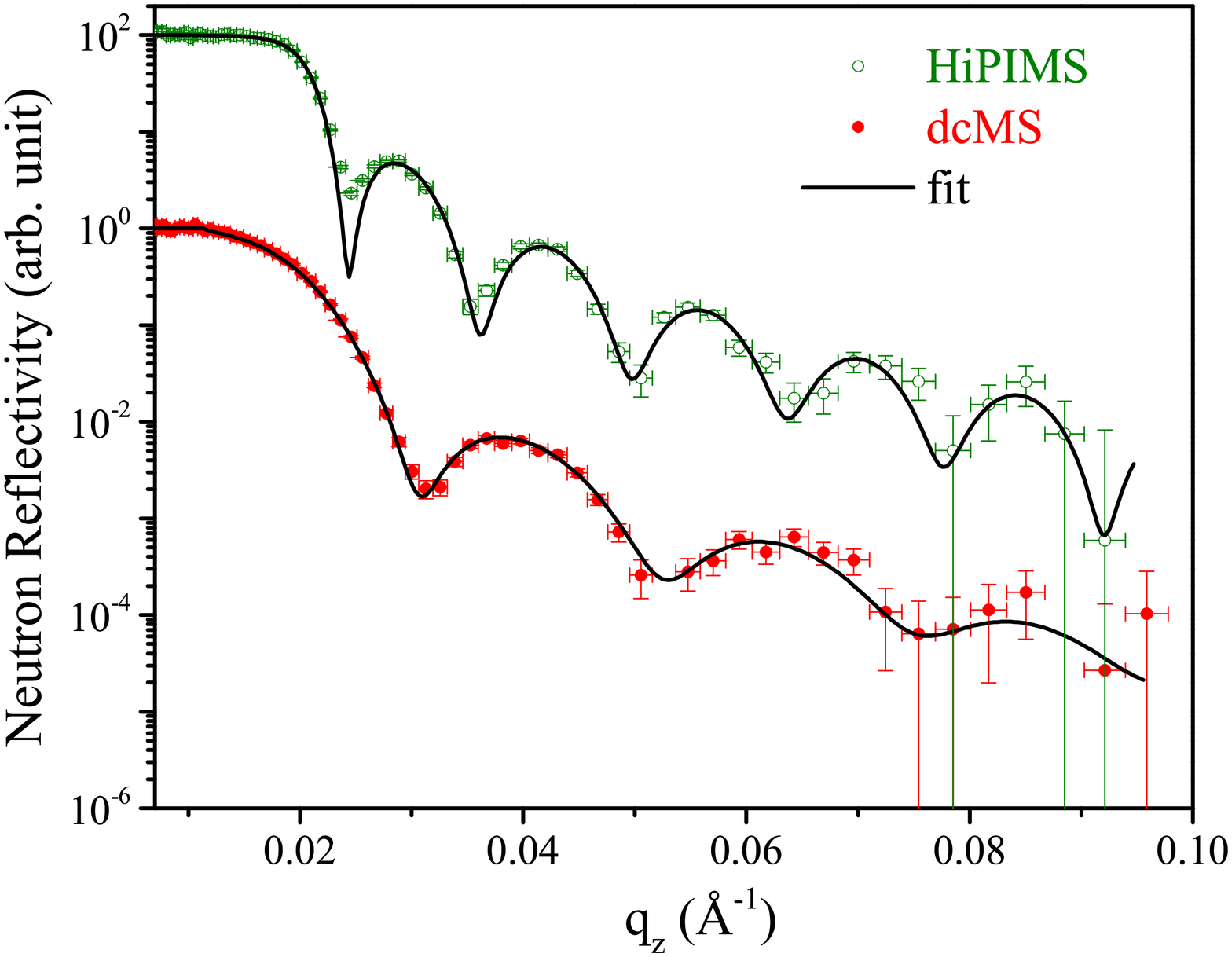} \vspace{-2mm}
\caption{\label{NR} (Color online) Neutron reflectivity of the
deposited carbon thin films using dcMS and HiPIMS.} \vspace{-1mm}
\end{figure}

\subsection{X-ray and Neutron Reflectivity}
\label{3.2}XRR is one of the well-known techniques for determining
thickness ($t$), roughness ($\sigma$) and number density ($\rho$)
or mass density of thin films. For x-rays, refractive index (n) is
a complex number and it is slightly less than unity, and can be
expressed as:
\begin{equation}\label{3.2.1}
    n=1-\delta-i\beta
\end{equation}
where,
\begin{equation}\label{3.2.3}
    \delta=\frac{\rho\lambda^{2}Zr_e}{2\pi}
    ~~~and~~~
    \beta=\frac{\rho\lambda\sigma_a}{4\pi}=\frac{\lambda\mu}{4\pi}
\end{equation}
here $\delta$ and $\beta$ are dispersive and absorptive part of
the refractive index, $\sigma_a$ is absorption coefficient, $\mu$
is linear absorption coefficient and $r_e$ is classical radii of
electron. The x-ray scattering length density (SLD) is frequently
used as: (i) Re(SLD)=$\rho$Z$r_{e}$ and (ii)
Im(SLD)=$\rho\sigma_a$. Since absorption of x-rays takes place in
material, it carries information related to presence of vacancy,
defects etc.

\begin{table*}\vspace{5mm} \caption{\label{table1} Parameters, film thickness ($t$), scattering length density (SLD),
surface roughness ($\sigma$) obtained from fitting of x-ray and
neutron reflectivity of $a$-C thin films deposited using dcMS and
HiPIMS. Mass density ($\rho_{\mathrm{m}}$) has been calculated
from real part of SLD (Re$_{\mathrm{SLD}}$).}

\begin{tabular}{|cc|ccc|ccccc|}
  \hline
  Technique/ &&&dcMS&&&&HiPIMS&&\\
  Parameter && XRR && NR && XRR && NR& \\
  \hline
  ($t$$\pm$5) \AA && 251 && 256 && 384 && 391& \\
  Re$_{\mathrm{SLD}}\times$10$^{-5}$\AA$^{-2}$ && 1.91$\pm$0.01 && 0.66$\pm$0.002 && 1.92$\pm$0.02 && 0.71$\pm$0.005& \\
  Im$_{\mathrm{SLD}}\times$10$^{-8}$\AA$^{-2}$ && 3.05 && 0 && 33 && 0& \\
  $\rho_{\mathrm{m}},$\,g/cm$^{3}$ && 2.25$\pm$0.01 && 1.98$\pm$0.004 && 2.26$\pm$0.02 && 2.14$\pm$0.005& \\
  ($\sigma$$\pm$2)\AA && 6.4 && 8 && 8.0 && 10& \\
    \hline
\end{tabular}
\end{table*}\vspace{1mm}

On the other hand, neutron reflectivity is also a frequently used
technique specially for magnetic thin films, as neutrons carries a
magnetic moment~\cite{OTT2007CRP}. In addition, NR has been
frequently used for low Z materials and polymers due to large
scattering cross-sections for H and D~\cite{PENFOLD2002COCIS}. In
view of this, it is surprising to note that NR has not yet been
used to study $a$-C thin films even though the mass density of C
and Si is very close. Since neutron scattering takes place from
nucleus, and therefore the neutron scattering length ($b$) can
differ significantly between neighboring elements. For C and Si,
$b$ is 6.5\,fm and 4.15\,fm, respectively. Similar to x-rays, the
refractive index for neutrons can be written as:
\begin{equation}\label{3.2.2}
    n\approx1-\frac{\rho\lambda^{2}b}{2\pi}
\end{equation}
Here $\rho$ is number density and the imaginary term has been
omitted as for C and Si it is about six orders of magnitude
smaller than dispersive part due to smaller absorption of
neutrons.

It may be noted that XRR has been frequently used to study $a$-C
films~\cite{Ferrari2000, PATSALAS20051241}. Generally, $a$-C thin
films are deposited on a Si or SiO$_{2}$ substrate. X-ray
scattering contrast between the substrate and C is rather poor due
to nearly equal number densities Re(SLD)$_{\mathrm{C}}$ =
1.87$\times$10$^{-5}$\AA$^{-2}$ (mass density
$\rho_{m}$=2.2\,g/cm$^{3}$) and Re(SLD)$_{\mathrm{Si}}$ =
1.98$\times$10$^{-5}$ \AA$^{-2}$ ($\rho_{m}$=2.3\,g/cm$^{3}$). In
the study of $a$-C films with XRR, apart from thickness ($t$) and
roughness ($\sigma$), the density of the film has been often
determined. Generally in an angle dispersive XRR measurement
typical errors in determination of density are about
5\%~\cite{WALLACE1995APL}. In addition, due to similar values of
Re(SLD) for C and Si, the critical angle is not well defined and
determination of density from XRR measurement could easily have
another 5\% error, signifying that even from a very careful
experiment, the density of $a$-C film could have errors exceeding
10\%.

On the other hand, neutron has a large scattering contrast between
C and Si due to large differences in their scattering lengths and
resulting number densities for neutrons are $\rho_{\mathrm{C}}$ =
7.33$\times$10$^{-6}$\AA$^{-2}$ and $\rho_{\mathrm{Si}}$ =
2.04$\times$10$^{-6}$ \AA$^{-2}$. In addition, absorption of
neutrons is almost negligible unlike x-rays and therefore, number
densities can be determined more precisely~\cite{WALLACE1995APL}.
Moreover, by doing neutron reflectivity measurements in the time
of flight (ToF) mode (similar to energy dispersive), the sample is
kept at a fixed angle of incidence and therefore illumination
remains constant. Hence, foot print effects can be avoided
completely and the critical angle is well-defined so that density
can be determined more accurately. Our comparison of both XRR and
NR, clearly shows the limitation of XRR determination of density
specially for carbon thin films.

XRR and NR patterns for both samples are shown in fig.~\ref{CXRR}
and ~\ref{NR}, respectively. They were fitted using a Parratt32
software package~\cite{lgparratt} to obtain $t$,
$\rho_{\mathrm{m}}$ and $\sigma$ and are shown in
table~\ref{table1}. As can be seen there, the value of thickness
and roughness obtained from XRR and NR are similar (within
experimental errors), as expected. But mass density of films
calculated from XRR and NR are at large variance and the value of
densities both for dcMS and HiPIMS deposited films are similar at
about 2.25\,g/cm$^{3}$. It may be noted that the mass density of
Si is about 2.3\,g/cm$^{3}$. Therefore, it may be possible that
rather than film, the substrate density is measured. On the other
hand, we find that the values of mass densities obtained from NR
measurements are 1.98 and 2.14\,g/cm$^{3}$, respectively for dcMS
and HiPIMS deposited films. Clearly, films deposited with HiPIMS
have larger density and it can only be probed by doing NR
measurements in the ToF mode.

An additional feature that can be seen from the XRR pattern is the
behavior of critical edges for the samples deposited with dcMS and
HiPIMS as shown in the inset of fig.~\ref{CXRR}). Such behavior
can only be fitted taking a much higher value for the imaginary
part of SLD. From table ~\ref{table1}, we find that the value of
Im$_{\mathrm{SLD}}$ is about 10 times higher for the film
deposited using HiPIMS. Such a significant increase in
Im$_{\mathrm{SLD}}$ can be understood by analyzing the XPS and XAS
results, which are presented in next sections.

\subsection{X-ray Photoelectron Spectroscopy} \label{3.3}

\begin{figure} \center \vspace{-1mm}
\includegraphics [width=90mm,height=75mm] {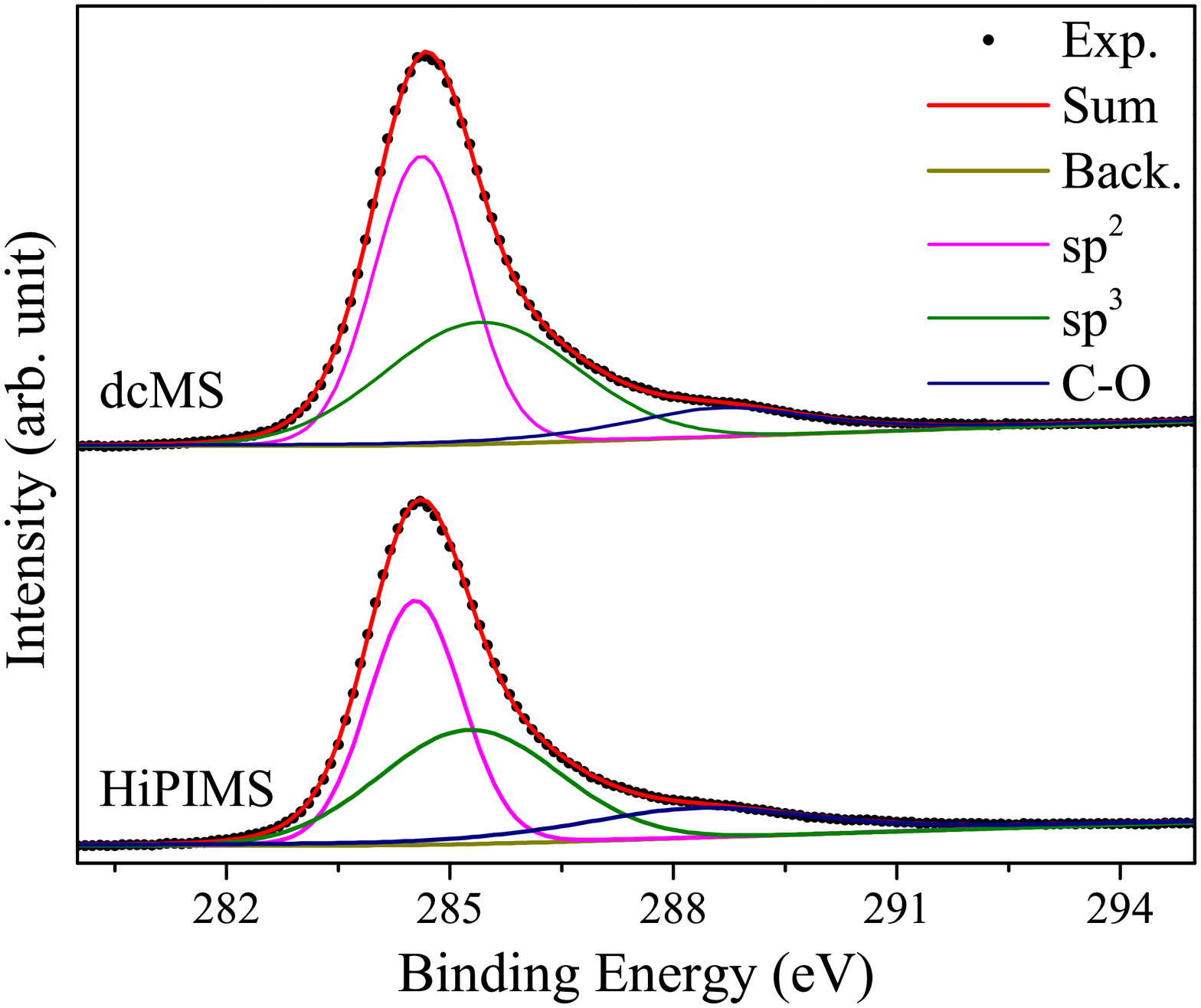} \vspace{-6mm}
\caption{\label{xps} (Color online) C 1s x-ray photoelectron
spectroscopy of \emph{a}-C thin films deposited using dcMS and
HiPIMS.} \vspace{-1mm}
\end{figure}

X-ray photoelectron spectroscopy (XPS) is one of the well-known
technique to determine sp$^{2}$ and sp$^{3}$ fractions in a $a$-C
sample. The XPS spectra shown in fig.~\ref{xps} can be
deconvoluted in three components (i) sp$^{2}$ (ii) sp$^{3}$ and
(iii) C bonded with oxygen, following
ref.~\cite{XPSPEAKFITPROCEDURE}. Using XPSPEAK41 peak fitting
software, we can find that experimental data matches well with the
calculation. Obtained fitting parameters are shown in
table~\ref{table2}. The sp$^{3}$/sp$^{2}$ hybridization ratio
found are 0.83 and 0.88, respectively for dcMS and HiPIMS
deposited films. This clearly shows that $a$-C film deposited
using HiPIMS have lower sp$^{2}$ fraction. It may also be noted
that the C-O fraction is also large in the $a$-C films deposited
using HiPIMS. Although XPS measurements clearly show differences
for the $a$-C thin films deposited using HiPIMS in terms of larger
sp$^{3}$/sp$^{2}$ hybridization ratio and C-O fractions, they can
be more precisely seen in the C K-edge spectra shown in the next
section.

\begin{table}\vspace{5mm}
\caption{\label{table2} Fitting parameters obtained form XPS data
of $a$-C thin films deposited using dcMS and HiPIMS techniques.}
\begin{tabular}{| c| c c | c c | }
  \hline
  Technique&~~~~~~~~dcMS&&~~~~~~~~HiPIMS&\\
  \hline
  Feature & BE (eV) & hyb.(\p) & BE(eV) & hyb.(\p) \\
  \hline
   sp$^{2}$ & 284.6 & 48.1 & 284.6 & 44.2 \\
   sp$^{3}$ & 285.4 & 39.7 & 285.3 & 38.7 \\
   C-O & 288.6 & 12.2 & 288.4 & 17.1 \\
  \hline
\end{tabular}
\end{table}

\subsection{X-ray Absorption Near Edge Spectroscopy} \label{3.4}
\begin{figure} \center \vspace{-1mm}
\includegraphics [width=90mm,height=75mm] {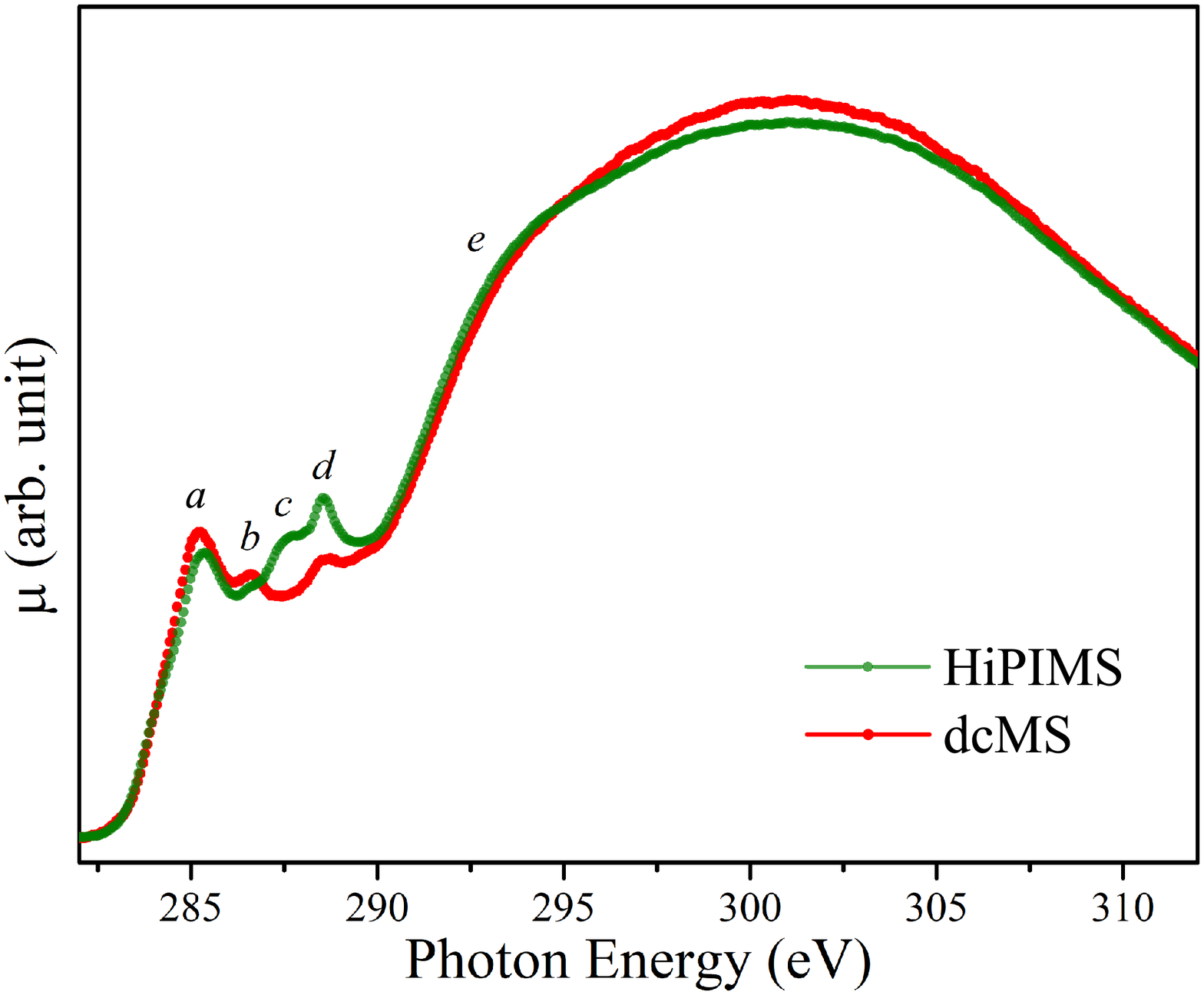} \vspace{-8mm}
\caption{\label{XAS1} (Color online) X-ray absorption near edge
spectra of the C-films were measured in TEY mode. Inset shows a
closer view of derivative of recorded data.} \vspace{-1mm}
\end{figure}

X-ray absorption near edge spectroscopy (XANES) is one of the best
technique to probe the local structure. C K-edge XANES spectra was
measured in total electron yield mode. Samples deposited using
dcMS and HiPIMS were measured under UHV conditions as shown in
fig.~\ref{XAS1}. Pre and post-edge normalization has been applied
using Athena software package~\cite{Ravel:ph5155}. The C K-edge
spectra have several features assigned as $a$, $b$, $c$, $d$ and
$e$. The feature \emph{a} (285.5\,eV) is due to
1s$\rightarrow\pi^{\star}$ transition (or $sp^{2}$ hybridization),
the faint feature \emph{b} is ambiguous and may be because of
hybridization of carbon with nitrogen or hydrogen when exposed to
atmosphere~\cite{BENNY2006SMALL}, features $c$ and $d$ are due to
hybridization of carbon with oxygen C-O and C=O, respectively and
$e$ is because of $sp^{3}$ hybridization. Comparing the films
deposited with dcMS and HiPIMS, following difference are
noteworthy (i) feature $a$ is more intense is dcMS deposited film
(ii) features $c$ and $d$ are more intense in HiPIMS deposited
film (iii) for HiPIMS deposited films we also find that the
feature $b$ is almost absent and the feature $e$ is somewhat more
prominent.

We find that the information obtained from C K-edge XANES
measurements is in agreement with other results obtained in this
work. Films deposited with HiPIMS have lower sp$^{2}$ fractions
and the most remarkable change can be seen in terms of
significantly stronger C=O and C-O features ($c$ and $d$). Since
in both cases films were prepared under similar deposition
conditions, it is unlikely that oxygen atoms get incorporated only
in HiPIMS deposition process. From the AFM measurements
(fig.~\ref{afm}) we can see that while dcMS deposited films have a
rather smooth texture, those deposited with HiPIMS have particles
about 5\,times larger in the form of hillocks that are surrounded
with a porous microstructure. Since the density of HiPIMS
deposited films (obtained from NR measurements) is larger despite
having a porous structure, it can be assumed that the hillocks
formed here are dense and have a large sp$^3$ fraction. Since in
the HiPIMS process there is an enhancement in the ions, these ions
not assist in nucleation of larger particles due to enhanced
mobility but they may also bombard the growing film and lead to
microstructure that is a combination of dense hillocks surrounded
by pores. When samples are exposed to atmosphere those pores get
filled with oxygen leading to a significant carbon oxygen bonding
as observed by both XPS and XANES. We believe that by further
tuning the deposition parameters in the HiPIMS process, the
formation pores can be avoided or increased. In some applications
in fact films with large porosity are required
~\cite{LAUSEVIC2013Carbon, SCHOPF2017Carbon}. Such types of films
have higher effective surface area that is desirable for high
energy devices. In either way HiPIMS process offers a possibly to
control the microstructure which is not generally possible in
typical dcMS processes.

\section{Conclusion}
In this work we studied $a$-C thin films deposited using dcMS and
HiPIMS processes. The microstructure, electronic properties and
the density of the films were measured. We show that while XRR
have limitations in measuring small differences in the density of
$a$-C films due to poor scattering contrast and sample alignment.
NR in ToF mode can be used for precise measurement of density of
carbon thin films. Our results on the micro, electronic and film
density correlate well and signify that HiPIMS process leads to
formation of dense particles but at the same time bombardment of
ions also increases pores. By further tuning the process it is
believed that even denser and high sp$^3$ fraction carbon thin
films can be obtained by HiPIMS.

\section*{Acknowledgments}
Authors would like to acknowledge Layanta Behera for help in
sample preparation, V. R. Reddy and Anil Gome for XRR, Niti and
Seema for NR, Rakesh Kumar Sah for XAS, Mohan Gangrade for AFM and
Prakash Behara for XPS measurement. We are thankful to A. K. Sinha
for support and encouragements. Authors thank the Department of
Science and Technology, India (SR/NM/Z-07/2015) for the financial
support and Jawaharlal Nehru Centre for Advanced Scientific
Research (JNCASR) for managing the project.

\section*{References}


%

\end{document}